\documentstyle[aps,pra,graphicx,twocolumn]{revtex}
\input epsf
\newcommand{\infig}[2]{\centerline{\epsfxsize=#2\columnwidth \epsfbox{#1}}\vspace{5mm}}
\begin{document}
\title{ Experimental verification of the Heisenberg uncertainty
principle\\
for hot fullerene molecules}
\author{Olaf Nairz, Markus Arndt, Anton Zeilinger}
\address{Universit\"at Wien, Institut f\"ur Experimentalphysik, Boltzmanngasse 5,
A-1090 Wien, Austria}
%
\maketitle
\begin{abstract}
The Heisenberg uncertainty principle for material
objects is an essential corner stone of quantum mechanics and
clearly visualizes the wave nature of matter. Here we report a
demonstration of the Heisenberg uncertainty principle for the most
massive, complex and hottest single object so far, the fullerene
molecule C$_{70}$ at a temperature of 900 K. We find a good
quantitative agreement with the theoretical expectation: $\Delta x
\times \Delta p = h$, where $\Delta x$ is the width of the
restricting slit, $\Delta p$ is the momentum transfer required to
deflect the fullerene to the first interference minimum and
h is Planck's quantum of action.
\end{abstract}
\pacs{03.65.-w,03.65.Ta,03.75.-b,39.20.+q}

Complementarity is one of the essential paradigms of quantum
mechanics \cite{BohrComplementarity}. Two quantities are mutually
complementary in that complete (or partial) knowledge of one
implies the complete (or partial) uncertainty about the other and
vice versa. The most generally known case is the complementarity
between position and momentum, as expressed quantitatively in the
Heisenberg uncertainty principle $\Delta x \times \Delta p \ge
\hbar /2$. For neutrons the uncertainty relation has been
demonstrated already back in 1966 by Shull
\cite{UncertaintyNeutron}. Following the growing experimental
efforts in atom optics during the last decade,  the uncertainty
principle has shown up implicitly in several experiments and has
also been explicitly investigated in both the spatial
\cite{UncertaintyAtomsSpatial} and in the time domain
\cite{UncertaintyAtomsTime}.

While being a physical phenomenon of interest in its own right, the complementarity between momentum and position
is also an important factor for practical purposes: for example it is
applied
for the preparation of transverse coherence in all experiments
using collimated beams, a fact that can be mathematically
phrased using the van Cittert-Zernike theorem \cite{Cittert,Zernike,BornWolf}.

There are good reasons to believe that complementarity and the
uncertainty relation will hold for all sufficiently well isolated
objects of the physical world and that these quantum properties
are generally only hidden by technical noise for larger objects.
It is therefore interesting to see how far this quantum mechanical
phenomenon can be experimentally extended to the macroscopic
domain.

Here we report on an experiment investigating for the first time in a quantitative way the
uncertainty relation upon diffraction at a single slit for a
molecule as complex, massive and hot as the fullerene C$_{70}$ ($m=840$ amu) at an internal and
translational temperature of 900 K.

It is well known that the limit $\hbar/2$ of the uncertainty
relation $\Delta x \times \Delta p \ge \hbar /2$ is only reached
for particular wave packets, for example of the Gaussian type. Evidently, the wave packet after
passage through a rectangular slit is very different from
this minimal uncertainty shape. This is also reflected in the far field
distribution which is described by the well known sinc-function
rather than a Gaussian. It is therefore a matter of definition and
convenience which quantities to take as a measure of the position and
momentum uncertainty in our case.  Obviously, for a wave traversing a slit,
one can take the slit width to be the measure of the spatial
uncertainty $\Delta x$. The momentum uncertainty $\Delta p$ can be
related to the angular spread due to diffraction at the slit.
Quantitatively, we define it as the momentum required to reach the
first order interference minimum, which for small angles, as used
in our experiment, lies in the direction $\theta \simeq \lambda
/\Delta x = \Delta p/p = \lambda  \Delta p/h $.
With this definition of the uncertainty we expect the following relation
\[ \Delta x \times  \Delta p = h
\]
to hold for the experiment.
It is this equality that we quantitatively test in the following for a
beam of C$_{70}$.

\begin{figure}[th!]
\infig{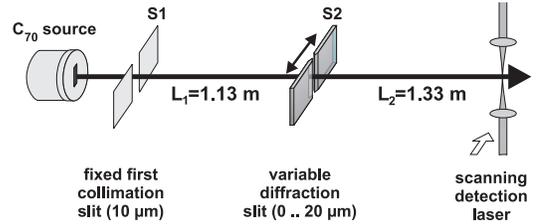} {.8} \caption{Setup of the experiment. A
thermal C$_{70}$ beam is produced by sublimation of fullerene
powder at 900 K. The beam is narrowed by S$_1$ and diffracted by
S$_2$. S$_1$ is fixed to $10 \,\mu m$. The width of slit S$_2$ is
varied with $\pm 30 \, nm$ accuracy for $\Delta x < 1\,\mu m$.\\ }
\end{figure}
In the actual experiment the minima are not very well defined due to the broad de Broglie wave
spectrum of our fullerene source. However, we can determine the
full width at half maximum (FWHM) values of the central lobe with good precision.
The far field diffraction pattern after a single
slit is given by $I(\theta) \sim (\sin(u)/u)^2$, where $u= \Delta x \cdot k
\theta/2$ and  $k=2 \pi/\lambda$ is the wave vector of the incident
fullerene molecule.
The half width at half maximum (HWHM) for the diffraction curve is then given by the solution of
$(\sin(u_{1/2})/u_{1/2})^2=1/2$, which yields
\[
u_{1/2}=0.443 \pi= \frac{k \Delta x}{2} \theta = \frac{\pi}{\lambda}
\cdot \Delta x \cdot \frac{\Delta p_{1/2}}{p} = \frac{\pi \Delta x \Delta
p_{1/2}}{h}.
\]
Practically, this means that we obtain the momentum uncertainty as defined
above from the measured FWHM via $\Delta p = \Delta p_{1/2}/0.89$.

The setup of the experiment, shown in fig.1, is similar to that described
in a previous publication\cite{Arndt1999}. An
effusive thermal fullerene beam is produced at
about 900 K. The velocity spread was as large as $\Delta v/v \sim 0.6$ and was taken into
account into the numerical description of the experiment.

The molecular beam is collimated by two piezo-controlled slits.
The width of the first slit $S_{1}$, is fixed at 10 $\mu$m,
while the width  $\Delta x$  of the second slit, $S_{2}$ -- which is located at the distance $L_1=113$ cm further
downstream -- can be varied to investigate the position-momentum uncertainty
relation.

In order to also quantitatively describe the  experiment the
properties of the slits have to be known rather precisely. The
slits (Piezosysteme Jena) are made of two silicon edges  mounted
on piezo-controlled  flexure stages. We obtain information
about the slit opening in three different ways: from the applied piezo voltage, from the reading
of a strain gauge mounted to the slits and finally from the total
number of molecules  passing through the slit at a given opening.
While the piezo voltage can be kept stable to better than $\Delta
U/U < 10^{-4} $ it is well known that piezos show creep,
hysteresis and non-linearities. However, it turned out in the
experiments that the passive stability over a typical time of 1
hour was of the order of 50 nm, as can be judged from the
stability of the diffraction patterns.  From a calibration of the
hysteresis curve we determine the change of the slit opening as a function of the piezo voltage. In
order to know the absolute slit width we determined the zero
position by measuring the  number of molecules passing through the slit, when
it was being closed. We estimate this method to be accurate to
within $\pm 30$ nm.

We extract the momentum spread $\Delta p$ after $S_{2}$ from
the FWHM of the detected molecular beam $\Delta X_{exp}$ in
the detection plane,  which is separated from $S_{2}$ by the
length $L_2=133$ cm. The observed distribution function
$f_{exp}(x)=D(x)\otimes M(x)$ is actually a convolution of the
detector resolution function $D(x)$ and the real molecular beam
profile $M(x)$.

The scanning laser ionization detector has been characterized in
depth in a previous publication \cite{Nairz}. For our present
experiments with C$_{70}$ the FWHM of the detector response was
determined to be $D=10\pm 0.5 \, \mu m$ at a laser power of P=10
W. The effective FWHM detector height at this power was measured
to be $\sim 1$ mm.

\begin{figure}[ht!]
\infig{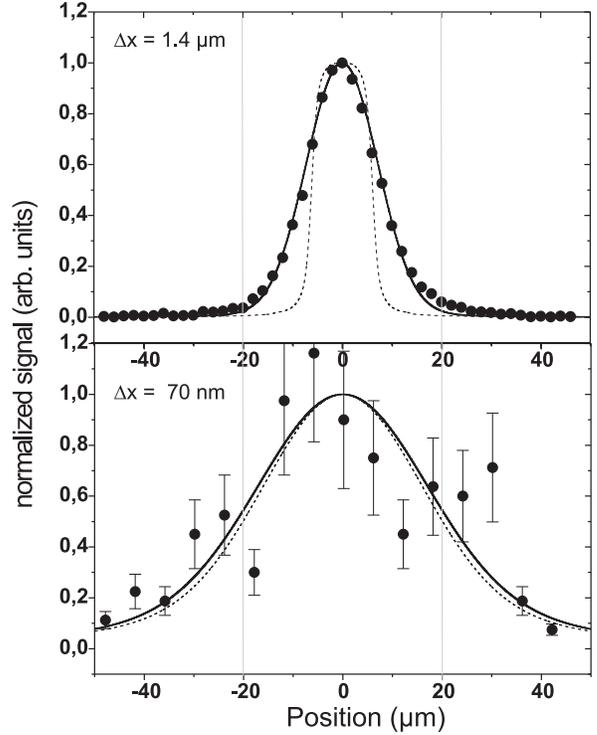} {.9}
\caption{Measured  molecule distribution in the detection plane after passing through a
piezo-controlled silicon slit having a width of $\Delta x = 1.4 \,\mu m$  (top) and  $\Delta x \sim 70 nm$ (bottom).
Both the  quantum mechanical calculation (continuous line) and the experiment (circle) show an
increase of the beam width when going from medium (top) to narrow (bottom) slit widths. The dotted line indicates the
wave calculation before the convolution with the known detector
profile.}
\end{figure}

The second contribution, related to the measured molecular beam
profile $M(x)$, is composed of both the classical collimation and
the momentum spread due to the quantum uncertainty. In order to
compare the experiment with the uncertainty relation derived
before, we concentrate in the following on the half-width values
of these components only. Since the classical FWHM shadow width  $\Delta
X_{cl}$ and the quantum contribution $\Delta X_{qu}$
are completely independent their influence can be added
quadratically to yield the FWHM value of $M(x)$, which we
denominate as $\Delta X_{mol}$.  The classical contribution
$\Delta X_{cl}$ can be derived from a simple geometrical
shadow model. Taking the measured and the classically expected
widths we can then deduce the contribution to the beam width due
to the quantum uncertainty and we finally relate this spatial
information to the corresponding momentum uncertainty, which then
reads:

\begin{equation}
\Delta p = \frac{p_z}{0.89 L_2}\left(\left[\left(\Delta X_{mol}\right)^2
-\left(\Delta X_{cl}\right)^2\right]^{1/2} -\Delta x \right),
\end{equation}
where $p_z$ is the longitudinal momentum of the molecule.

To trace out the uncertainty relation we varied the width of the
second slit from about $20\,  \mu m$ down to roughly $50 \, nm$ and
record the molecular beam width in the detection plane.
In figure 2 we show the measured molecular beam profiles as full circles for two different
widths of the second collimation slit. We see a relatively narrow beam of
$ \Delta X_{exp}= 17 \, \mu m$ for  the slit width $\Delta x = 1.4\, \mu$m  (fig 2a) and
again a strong growth to  $\Delta X_{exp} =43\, \mu m$
for the slit width  $\Delta x = (0.07 \,\pm 0.03\,) \mu m$ (fig 2b).
The error bars in fig. 2 represent the statistical uncertainty due
to the very low count rate, in particular at the smallest slit
width.

The dashed line follows a full wave
calculation as described below in order to show the molecular
beam profile as given by  diffraction alone. The continuous curves represent the
same model but convoluted with the detector profile.

Close inspection of the data shows a good agreement between the
convoluted wave model and the experimental data.
This good agreement is the first demonstration
of single slit diffraction for a molecule as heavy, complex and hot as C$_{70}$.

Fig. 2b is actually an interesting
complement to the high contrast interference fringes of fullerenes
after diffraction at a nanofabricated grating with a grating
constant of 100 nm, which we could demonstrate in a previous
publication \cite{Arndt2001}.
The single slit pattern shown here is the envelope of the far-field grating
interference pattern. This provides a striking
proof of the wave nature of the fullerene
C$_{70}$ because it demonstrates that the previous minima must
have been due to destructive interference.
\begin{figure}[ht!]
\infig{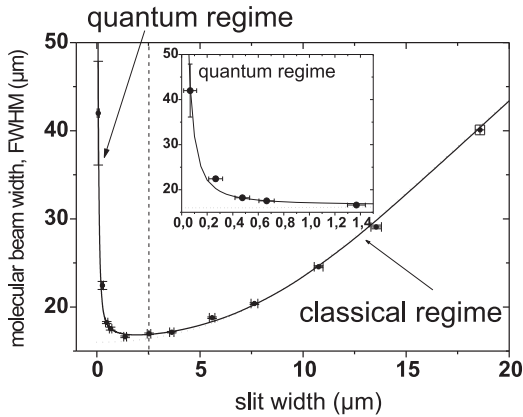} {.8} \caption{The experimental molecular
beam width (full circle) is compared with the quantum prediction
(continuous line) as a function of the slit opening $\Delta x$.
The agreement is excellent across the whole range of slit openings
($70 nm ... 20 \, \mu m$). A purely classical shadow model
predicts the dotted line and is in marked disagreement with the
data for $\Delta x < 4\mu$m. The latter is therefore designated as
the quantum regime and magnified in the inset of fig 3. }
\end{figure}
From the whole series of experiments with varying slit widths we have
extracted the FWHM values from the experiment and we
compare them with a quantum wave model in fig. 3.
An excellent agreement between expectation and experiment is found
throughout the whole range of values.
We can distinguish essentially two different regimes
corresponding to a  pure quantum regime (left part of fig. 3)
and a range which can be very well described using a classical ball
model (right part of fig. 3).
The continuous wave calculation curve and the dotted
classical line coincide almost completely down to a slit
width of about $\Delta x = 4 \,\mu m$. Below this value the quantum mechanical
momentum spread $\Delta p$ contributes significantly to the  beam width in the detection plane.
This quantum range is magnified in the inset of fig. 3.

The horizontal error bars in this picture have two components,
namely the precision both of the absolute zero and of the scaling
of the piezo translation as a function of the applied voltage.
Both are only important for a small slit width $\Delta x$. The
absolute zero (closed) position of the piezo slits is known with
an error of $\pm$ 30 nm, as mentioned above. The scaling with the
applied piezo voltage is non-linear and follows a hysteresis curve, which
has been calibrated. We estimate an uncertainty of $\pm 3\% $ in the
calibration of the hysteresis curve.

The vertical error bars estimate the uncertainty of the measured
width of the beam in the detection plane. For small slit widths
$\Delta x$ these values are obtained from a least squares Gaussian
fit to the detected curve. For large $\Delta x$ the marked
trapezoidal shape as well as the high signal-to-noise ratio permit
a direct reading of the experimental and theoretical FWHM values
with very high accuracy.

The  numerical simulations in fig. 3 are based on the fact that
the Schr\"{o}dinger equation of our time independent problem  is
formally equivalent to the Helmholtz equation and can therefore be
treated using all the methods well known from optics. The solution
is done in close analogy to the numerical approach as used in
\cite{Zeilinger} for neutrons and similarly in
\cite{Turchette,Carnal} for atoms. There is no free parameter in
the calculation except for a broadening of the detector resolution
by $3.5 \mu m$ with respect to the best detector resolution curves
recorded some time earlier \cite{Nairz}. This offset is most
likely explained by a residual tilt of 2.7 mrad between laser and
diffraction slit. This is in agreement with diffraction curves not
shown here, which were recorded using the same setup but  at half
the width of the first collimation slit.

Since in previous papers it has been pointed out that the form
factor of single slit diffraction may be influenced by the van der
Waals interaction between the molecule and the slit walls
\cite{Hegerfeldt,Arndt1999} one may wonder whether this effect may
become visible in the present experiment. However, the slit widths
here, except for the smallest, are much larger than in the former
grating diffraction experiments, where the effective slit width
was reduced by about 15 nm. Since the van der Waals potential above
a surface decreases with the third power of the object-wall
distance the effect becomes small for the present study although
the slit thickness is bigger than that of the previously used
SiN$_x$ gratings. For the smallest slit width, $\Delta x \sim 70 nm$, a
possible contribution is masked by the experimental error bar.

In the following we compare our findings with the
Heisenberg uncertainty relation between position and momentum. For
this we use the method as indicated further above: From the
measured beam width we separate the influence of the detector
resolution in a deconvolution procedure. The remaining molecular
beam width is then decomposed into its classical and quantum part.

\begin{figure}[ht!]
\infig{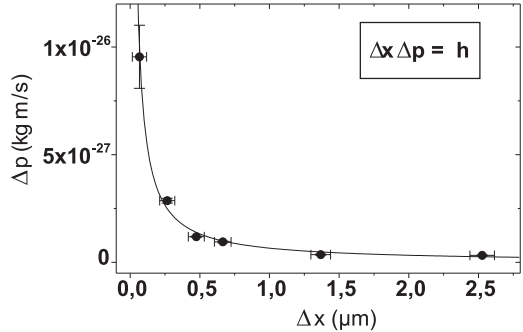} {.8} \caption{Experimental verification of
the uncertainty relation for C$_{70}$. The momentum values are
derived from the far field molecular beam widths as described in
the text. The position uncertainty is given by the width of the
second slit. The experiment is in very good agreement with the
wave model expectation: $\Delta x \times \Delta p = h$ and in
marked disagreement with a purely classical shadow model.}
\end{figure}
We can then plot $\Delta p$ as derived from eqn. 1 as a function
of $\Delta x$ for slit openings lying well in the quantum regime
and obtain fig. 4. The full circles represent the values extracted
from the experiment with error bars directly related to those of
the inset of fig. 3. The continuous line follows our original
expectation function $\Delta x \times \Delta p =  h$
without any additional fit parameter. We regard the good quantitative
agreement between the data points and the predicted
curve as a good support for the validity of the
Heisenberg uncertainty principle for the fullerene C$_{70}$ - i.e. in a
complexity region not studied so far.

\begin{acknowledgments}
We acknowledge help in the setup of the experiment by Julian Voss-Andreae,
Claudia Keller,  Gerbrand van der Zouw and Julia
Petschinka.
This work has been supported by the European TMR network, contract no.
ERBFMRXCT960002 and by the Austrian Science Foundation (FWF), within
the project F1505. O.N. acknowledges a scholarship from the Austrian Academy of Sciences.
\end{acknowledgments}

\end{document}